# Detection of Children Abuse by Voice and Audio Classification by Short-Time Fourier Transform Machine Learning implemented on Nvidia Edge GPU device

Jiuqi Yan[1], Yingxian Chen[1], W.W.T. Fok[1]
[1] Electrical and Electronic Engineering, The University of Hong Kong, Hong Kong

*Corresponding author's email: yjqhku@connect.hku.hk

**Abstract.**
The safety of children in children home has become an increasing social concern, and the purpose of this experiment is to use machine learning applied to detect the scenarios of child abuse to increase the safety of children. This experiment uses machine learning to classify and recognize a child's voice and predict whether the current sound made by the child is crying, screaming or laughing. If a child is found to be crying or screaming, an alert is immediately sent to the relevant personnel so that they can perceive what the child may be experiencing in a surveillance blind spot and respond in a timely manner. Together with a hybrid use of video image classification, the accuracy of child abuse detection can be significantly increased. This greatly reduces the likelihood that a child will receive violent abuse in the nursery and allows personnel to stop an imminent or incipient child abuse incident in time. The datasets collected from this experiment is entirely from sounds recorded on site at the children home, including crying, laughing, screaming sound and background noises. These sound files are transformed into spectrograms using Short-Time Fourier Transform, and then these image data are imported into a CNN neural network for classification, and the final trained model can achieve an accuracy of about 92% for sound detection.

## 1. Introduction

Children in school or children home are encountering risk of home accident, fighting among children and being abused. There were a few incidents that children in the children home in Hong Kong were abused by their caretaker and police investigation is required. The government called for using latest Artificial Intelligent to strengthen the monitoring and supervision of children home and generate real-time alerts to supervisors if there is any violent or abnormal behavior detected. Apart from using CCTV video image, the sound in the children home could also provide signal to reflect the situation happening in the children home. This research project develop algorithm and build a light-weight AI model for the analysis of audio wave form and classify the type of sound to assist the classification of the caretakers and children's behavior.

## 2. Related works

### 2.1 Machine Learning
Machine learning has achieved great success in the past decades in the direction of image classification[12], face recognition, unmanned autonomous vehicles, speech recognition, etc[3].Linnaeinma[4] conceived and proposed the model of BP neural, i.e. - the inverse model of automatic differentiation model. However, it did not create an academic wave at that time, but stagnated for a decade, and the backpropagation algorithm described in detail The backpropagation algorithm described in detail was introduced by Weibos[5]. After a few years, there have been algorithms combined with training, and many scholars involved in the field of neural network research at that time proposed the idea of combining MLP and BP training[6][7].

### 2.2 Neural Networks
Neural networks are an important machine learning technique with an overall structure similar to the nerves of the human brain and are designed to implement human brain functions. The most famous convolutional neural network in deep learning was proposed by Lecun et al[8]. It was the first true multilayer structural learning algorithm that uses spatial relativity to reduce the number of parameters to improve training performance. Based on the original multilayer neural network, a feature learning part was added, which mimics the human brain's hierarchy on signal processing.

## 2.3 Short-Time Fourier Transform

The Fourier transform only reflects the characteristics of the signal in the frequency domain and cannot analyze the signal in the time domain. In order to link the time domain and frequency domain, Gabor proposed the short-time Fourier transform (STFT)[2]. The process of STFT is: multiplying a time-limited window function h(t) before the Fourier transform of the signal, and assuming that the non-stationary signal is stationary within a short time interval of the analysis window, a set of local "spectrum" of the signal is obtained by shifting the window function h(t) on the time axis and analyzing the signal segment by segment.

## 2.4 The combination of STFT and CNN

By combining Short Time Fourier Transform (STFT) with Convolutional Neural Network (CNN), this project provides an advanced approach to sound signal processing and classification for preventing and solving child abuse in kindergartens. This combination of techniques allows for better characterization of children's voices and provides a powerful tool for real-time monitoring and protection of children's safety.

## 3. Our approach
### 3.1. Formatting of The Datasets

After collecting a certain amount of datasets in a children home, the datasets was divided in different directories. Dividing them into directories could allow easy loading of data using keras.utils.audio_dataset_from_directory. Before loading the data, the format of the data needs to be standardized and this data needs to be trimmed. This project uses Audacity to process the data to the nearest second, 48 kHz, so that the data can be batched in an effective way.

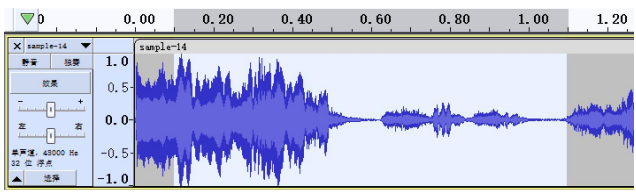

Figure 3.1 Formatting of the voice data

### 3.2. Conversion of signal domain

After this process, the datasets then contained batches of audio clips and integer tags. The audio clips were in the shape of (batch, sample, channel). The datasets contained only mono audio. The tf.squeeze function is used to remove the extra axes.

Finally the waveforms of the sound data were plotted, but these waveforms were represented in the time domain. This project uses Short Time Fourier Transform (STFT) to transform the waveform from a time domain signal to a frequency domain signal, converting the waveform plot to a spectrogram showing the change in frequency over time. The final import of the neural network model training is the spectrogram.

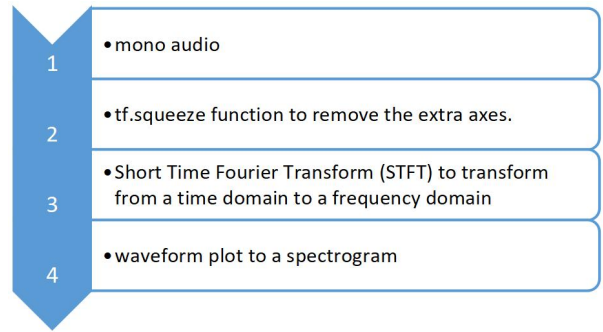

### 2.3. Short-Time Fourier Transform(STFT)

By moving the window function h(t) on the time axis, the signal is analyzed segment by segment to obtain a local "spectrum" of the signal. The short-time Fourier transform of a signal is defined as

$$\text{STFT}(t, f) = \int_{-\infty}^{\infty} x(\tau)h(\tau - t)e^{-j2\pi f\tau}d\tau$$

For a given time t, this can be considered as the spectrum at that moment. In particular, when the window function is taken, then the short-time Fourier transform degenerates to the conventional Fourier transform. To get the optimal localization performance, the width of the window function in the time-frequency analysis should be adjusted according to the signal characteristics, i.e., large window width for sinusoidal-type signals and small window width for impulsive-type signals.

### 3.4. Using convolutional neural network

After transforming all the sound signals into spectral images, a convolutional neural network will be used to process the data set.

The input to the model is a three-dimensional tensor with the shape (124, 129, 1). The first dimension represents the height, the second the width and the third the number of channels (in this case 1 channel). The model is a sequential model, meaning that the layers are stacked together in a sequential manner. There are 10 layers in the model and these layers are:

1. The first layer is the Resizing layer, this layer is responsible for resizing the input, resizing the input data from the original (124, 129, 1) to (32, 32, 1).
2. The second layer is the Normalization layer, which normalizes the input data to ensure that the data has a similar scale and range during training.
3. The third layer is the Conv2D layer, which is a convolutional layer that performs convolutional operations on the input using 32 3*3 convolutional kernels, and it extracts the features in the input data by convolutional operations.
4. The fourth layer is the Conv2D_1 layer, which is another convolutional layer that uses 64 3*3 convolutional kernels to perform convolutional operations on the output of the previous layer, which further extracts features and increases the complexity of the network.
5. The fifth layer is the MaxPooling2D layer, which is a maximum pooling layer that performs downsampling by selecting the maximum value in a 2*2 window. It reduces the size of the feature map while preserving the important feature information.
6. The sixth layer is the Dropout layer, which is used to randomly discard the output of some neurons to reduce the risk of overfitting. During the training process, the output of certain neurons will be set to zero with a certain probability.
7. The seventh layer is the Flatten layer, which spreads the output of the previous layer into a one-dimensional vector in order to feed it into the subsequent fully connected layers.
8. The eighth layer is the Dense layer, which is a fully connected layer with 128 neurons. It learns higher-level feature representations by connecting all neurons of the previous layer.
9. The ninth layer is Dropout_1 layer, which is another Dropout layer used to randomly discard the output of some neurons during training.
10. Finally the tenth layer is the Dense_1 layer, which is the final fully connected layer with 8 neurons, corresponding to the number of output categories of the model. It outputs the final classification results.

### *3.5. The ReLU function*

The ReLU[9] function is represented as follow：

ReLU (Rectified Linear Unit)：ReLU(x) = max(0, x)

The ReLU function has sparse activation on positive intervals relative to other activation functions (e.g., sigmoid and tanh). This means that when the input is negative, the output of ReLU is zero and the activated neurons will be suppressed. This sparse activation helps the sparse representation of the model and improves the expressiveness of the model. the ReLU function is linear on positive intervals and the gradient is always 1 when the input is positive. in contrast, the sigmoid and tanh functions have a gradient close to zero as the input approaches the saturation region, leading to the gradient vanishing problem. the linear activation of ReLU helps alleviate the gradient vanishing problem, allowing the The ReLU function is very simple to compute and only needs to compare whether the input is greater than zero and retain positive values. Compared with functions such as sigmoid and tanh, ReLU is faster to compute, especially when dealing with large-scale data, and can improve the efficiency of training and inference. the gradient of the ReLU function is always 1 on the positive interval, and there is no gradient explosion problem. This contributes to the stability of the training process and increases the convergence speed of the model.

### *3.6. Adaptive Moment Estimation optimizer*

The Adam (Adaptive Moment Estimation) optimizer is chosen for this experiment. The advantages of the Adam optimizer include: Adaptive learning rate: Adam dynamically adjusts the learning rate based on the historical gradients of the parameters. For different parameters and gradients with different time steps, it can provide the appropriate learning rate for each parameter, thus better adapting to the characteristics of the data; fast convergence: Adam can use a large learning rate in the initial training phase, and as training proceeds, the adaptive mechanism makes the learning rate gradually decrease, which helps speed up the convergence; friendly to sparse gradients: Adam is better at handling sparse gradients It can effectively handle sparse data commonly found in natural language processing and other fields; no need to manually adjust the learning rate: compared with some traditional optimization algorithms, Adam does not need to manually adjust the parameters of the learning rate, reducing the complexity of selecting and adjusting hyperparameters.

### *3.7. Implementation details*
This experiment was done on a laptop containing a Nvidia GTX1070. The program is done using the Tensorflow-GPU. The learning rate is 0.0001 and the epochs are 10. The optimizer is Adam, the evaluation metric is accuracy, and the loss function is cross-entropy loss function.

### 4. Result and Discussion
In this research, a simple convolutional neural network is used to classify and recognize the different voices of children recorded in the children home scene.

First use STFT to transform the sound waveform graph from a time domain waveform to a frequency domain waveform.

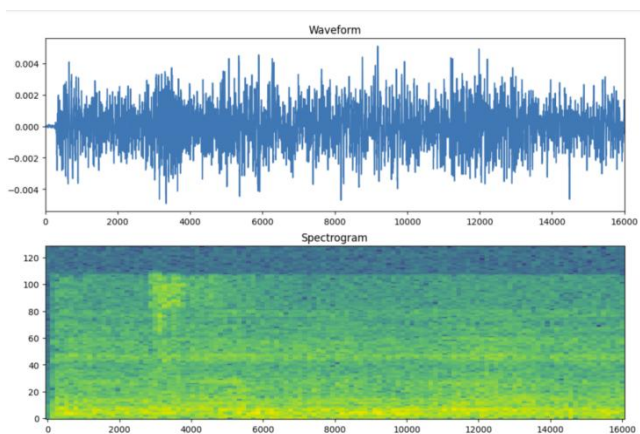

Figure 4.1 Short-Time Fourier Transform

The spectrograms are then imported into the CNN
From Figure 3.2, it can be observed that during the training process, val_loss gradually decreases from 0.9172 at the beginning to 0.3680, val_accuracy gradually increases from 0.4474 at the beginning to 0.8947, but decreases to 0.8421 after the last round of training, and val_accuracy also partially floats during the increase.

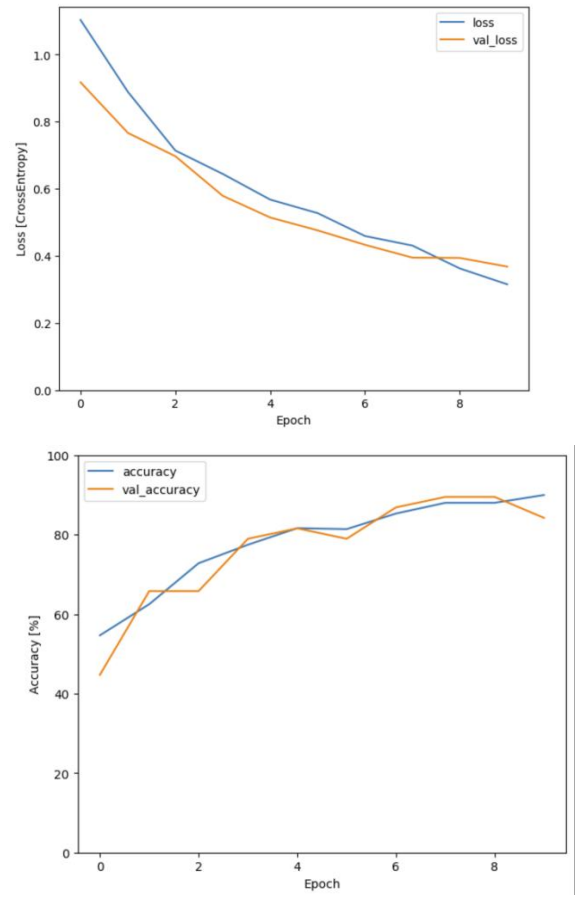

Figure 4.2 Experiment result

Use the confusion matrix to check how well the model classifies each command in the test set

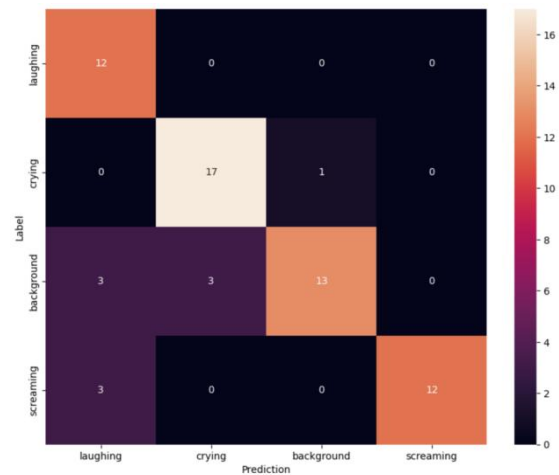

Figure 4.3 Confusion matrix

From the confusion matrix in Figure 4.3 we can see that the detection accuracy of the four Labels of the confusion matrix: laughing, crying, screaming and background sounds. laughing in this label, it is found that a small part of the data is misclassified as background sounds and In this label, three data were misclassified as bachground sounds. in this label, all data were correctly judged. In the label of background sounds, only one data was misclassified as crying, thus it can be seen that the accuracy of the model is higher for screaming due to the high recognition. The laughing and screaming sounds of children are similar to some extent, so it is easy to misjudge laughing as screaming.

Finally, run the inference on the audio file and import an audio file, in this case a child's crying sound is selected. After the file is imported, we check the accuracy of the model's prediction results.

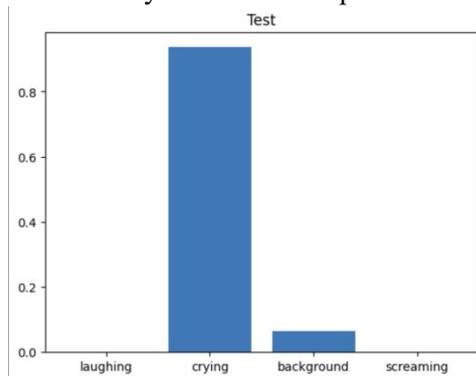

Figure 4.4 Prediction results

From Figure 3.4, it can be seen that after importing the child's crying sound, the model detects that the sound made by the child belongs to the classification 'Crying' and only about 0.08 probability belongs to 'Background' i.e. noise. This proves that the accuracy of the model reaches about 92%.

## 5. Conclusion

A combination of short-time Fourier transform (STFT) and convolutional neural network could be used for the classification of voice and sound in a children home. The STFT is used to turn the sound signals into spectrograms, and then the convolutional neural network is used to classify and predict the images. It is measured that the accuracy of the model for the detection of the child's voice remains above 90%, and basically it will not be misjudged as other voices.